\documentclass[journal=jacsat,manuscript=article]{achemso}

\usepackage[version=3]{mhchem} 

\title
{Device Applications of Heterogeneously Integrated Strain-Switched Ferrimagnets/Topological Insulator/Piezoelectric Stacks}

\author{Supriyo Bandyopadhyay}
\affiliation{Department of Electrical and Computer Engineering, Virginia Commonwealth University, Richmond, VA 23284, USA}
\email{sbandy@vcu.edu}

\keywords{Topological insulator, straintronics, magnetic anisotropy, strain-switched ferrimagnets}

\begin{document}


\begin{abstract}
  A family of ferrimagnets (CoV$_2$O$_4$, GdCo, TbCo) exhibits out--of--plane magnetic anisotropy when strained compressively and in--plane magnetic anisotropy when strained expansively (or vice versa). If such a ferrimagnetic thin film is placed on top of a topological insulator (TI) thin film and its magnetic anisotropy is modulated with strain, then interfacial exchange coupling between the ferrimagnet (FM) and the underlying TI will modulate the surface current flowing through the latter. If the strain is varied continuously, the current will also vary continuously and if the strain alternates in time, the current will also alternate with the frequency of the strain modulation, as long as the frequency is not so high that the period is smaller than the switching time of the FM. If the strain is generated with a gate voltage by integrating a piezoelectric underneath the FM/TI stack, then that can implement a transconductance amplifier or a synapse for neuromorphic computation.
\end{abstract}

\section{Introduction}

Heterostructures of strained ferromagnets/ferrimagnets and topological insulators are interesting constructs. 
Consider a ferromagnetic or ferrimagnetic layer (FM) deposited on a topological insulator (TI) thin film. The magnetic anisotropy of the FM layer can be changed from out-of-plane to in-plane, or vice versa, with strain \cite{kim,kim1,vaz,el-khabchi,poon}.  Interfacial exchange interaction between the spins in the magnet and the surface electrons in the TI can then open or close a bandgap in the TI as the magnetic anisotropy changes (or the easy axis changes) because of the breaking of time reversal symmetry. This modulates the surface current through the TI, turning it on and off \cite{kim1,ghosh} to implement a binary ``switch''. It has been shown that this switching action can take place in less than a nanosecond in some cases, with a conductance ON/OFF ratio of $\sim$100:1 and with a steep sub-threshold slope smaller than the Boltzmann limit of 60 mV/decade \cite{kim1}.

The binary switch described above is of course a digital device. Instead of reversing the strain {\it abruptly} from compressive to tensile, or vice versa, to realize a binary switch, we can vary the strain {\it gradually} from compressive to tensile or tensile to compressive. This will make the magnetic anisotropy change continuously with time and  the surface current through the TI will change in an {\it analog} fashion. For device implementation, we can generate the strain with a gate voltage if we integrate a poled piezoelectric layer underneath the FM/TI stack and apply the gate voltage across its thickness. The voltage will generate a strain in the piezoelectric which will be transferred to the FM through the ultrathin TI layer.  Compressive strain is generated by a voltage whose polarity is opposite to the direction of poling of the piezoelectric layer and tensile strain is generated by a voltage of the opposite polarity. Thus, by changing the gate voltage continuously from positive to negative, we can change the surface current in the TI from a positive peak to the negative peak (or vice versa)  continuously and this implements a transconductance amplifier, where the gate voltage is the input and the surface current is the output.  

If instead of a time varying gate voltage, we apply a static gate voltage, the current through the TI will be fixed (in time), but we can vary its magnitude with the gate voltage. This will implement a {\it synapse} for neuromorphic computation. Thus, the FM/TI/piezoelectric stack can have multiple device applications.

\section{Device design}

\begin{figure*}[h]
    \begin{center}
        \includegraphics[width=0.99\textwidth]{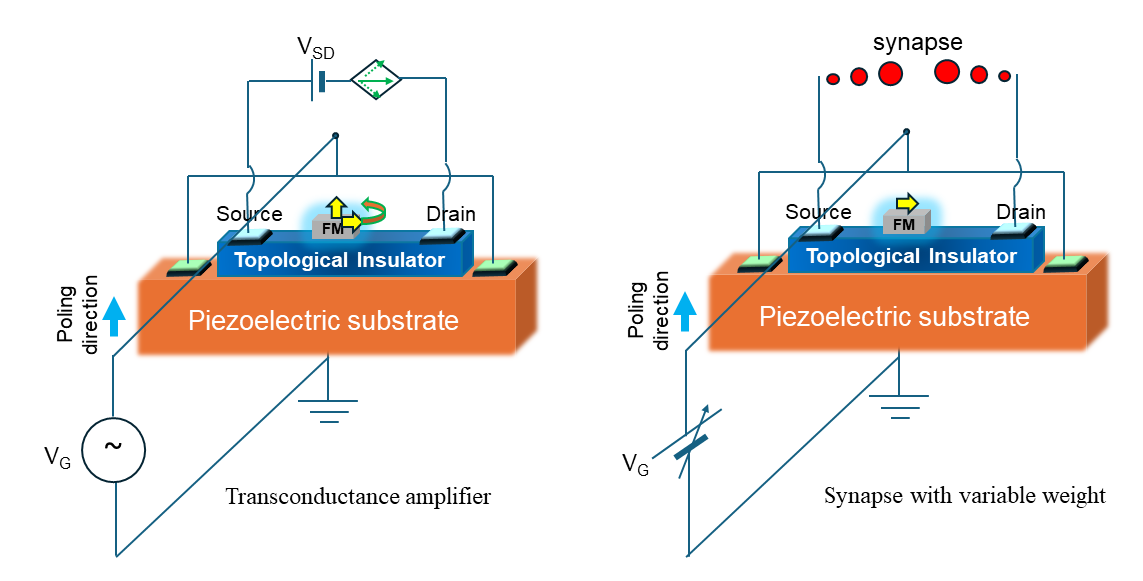}
    \end{center}
    \caption{(a) The proposed transconductance amplifier built with a piezoelectric/topological insulator/ferromagnet (or ferrimagnet) stack, (b) A synapse built with the same stack.}
    \label{fig:Device}
\end{figure*}

The device schematic for this idea is shown in Fig. \ref{fig:Device}. A thin layer (5-10 nm) of a TI is deposited on top of a poled piezoelectric substrate. A  layer of the FM is deposited on top of the TI and fashioned into a mesa. 

Two contact pads, delineated on the piezoelectric's surface, flank the FM and are shorted together. Application of a positive voltage $V_G$ between the two shorted contacts and the ground will generate a compressive strain in the piezoelectric because the resulting electric field is opposite to the direction of poling \cite{cui}. The strain is actually biaxial -- compressive along the line joining the two contacts and tensile along the direction perpendicular to that line \cite{cui}. Gate voltage of the opposite polarity will reverse the signs of the strains. The strain will be transferred through the ultrathin TI layer to the FM and switch its magnetic anisotropy from in--plane to out--of--plane, or vice versa, depending on the voltage polarity   \cite{kim,el-khabchi}. That will modulate the current flowing through the TI \cite{kim,kim1,ghosh} between the ``source'' and ``drain'' contacts.  As the gate voltage swings from positive to negative, the current through the TI (source--to--drain current) will also alternate its polarity synchronously with the gate voltage to implement a transconductance amplifier. To obtain the actual characteristics of the transconductance amplifier, one will have to solve the Landau-Lifshitz-Gilbert-Langevin equation in the FM and model transport in the TI using, for example, the non-equilibrium Green's function method. This is left for future work.

For implementing a synapse, we will apply a fixed gate voltage to generate a fixed source--to--drain resistance (see Fig. \ref{fig:Device}(b)). This resistance can be varied with the gate voltage to vary the synaptic weight.

\section{Discussion}

Generally speaking, a ferrimagnet-based construct will be different from the ferromagnet-based construct  studied in ref. [\citenum{kim1}] in an important way. In ref. [\citenum{kim1}], which used a ferromagnet instead of a ferrimagnet, the current polarity does not change if the gate voltage polarity is changed. Consequently, the device of ref. [\citenum{kim1}] cannot act as a true transconductance amplifier since the output current will be {\it rectified} when the gate voltage is sinusoidal. This does not happen if ferrimagnets are used. However, not all ferrimagnets are equal.  Some like CoV$_2$O$_2$ are ferrimagnetic only at low temperatures (below 150 K) and have weak saturation magnetizations, while others (GdCo, TbCo) do not suffer from these drawbacks.

A major advantage of this paradigm for realizing a transconductance amplifier or synapse is the energy efficiency. Very little gate voltage (few mV) is usually required to generate sufficient strain \cite{book,APR,JAP} to alter the magnetic anisotropy and thus alter the surface current in a TI. This will also result in a large transconductance which is very desirable in analog circuits.

Finally, quantum materials like topological insulators, Weyl semi-metals, etc. have been studied for more than a decade now with very few device proposals based on them. The recent flurry of device proposals \cite{kim1,ghosh,TI-transistor,WSM-transistor} is beginning to alter this landscape and bringing quantum materials to the forefront of applied physics and engineering.


\bigskip

{\noindent \bf Acknowledgments}
This work is supported by the US National Science Foundation under grant CCF-2504228. Discussions with Professors Avik Ghosh and Joseph Poon of the University of Virginia at Charlottesville are acknowledged with gratitude.

\end{document}